\documentclass[twocolumn,aps,showpacs]{revtex4}
\usepackage{graphicx}
\usepackage{dcolumn}
\usepackage{bm}
\begin{document}
\preprint{}
\title{$K^{\star}(892)^{0}$ Production in Relativistic Heavy Ion Collisions at $\sqrt{s_{_{NN}}}=130$ GeV}
\author{
\begin{flushleft}
C.~Adler$^{11}$, Z.~Ahammed$^{23}$, C.~Allgower$^{12}$, J.~Amonett$^{14}$,
B.D.~Anderson$^{14}$, M.~Anderson$^5$, G.S.~Averichev$^{9}$, 
J.~Balewski$^{12}$, O.~Barannikova$^{9,23}$, L.S.~Barnby$^{14}$
% ...
\end{flushleft}
\begin{center}(STAR Collaboration)\end{center}
}
% ...
\affiliation{$^5$University of California, Davis, California 95616}
% ...
\affiliation{$^{9}$Laboratory for High Energy (JINR), Dubna, Russia}
% ...
\affiliation{$^{11}$University of Frankfurt, Frankfurt, Germany}
\affiliation{$^{12}$Indiana University, Bloomington, Indiana 47408}
% ...
\affiliation{$^{14}$Kent State University, Kent, Ohio 44242}
% ...
\affiliation{$^{23}$Purdue University, West Lafayette, Indiana 47907}
% ...

\date{\today}
\begin{abstract}
We report the first observation of 
$K^{\star}(892)^{0}\rightarrow\pi K$ in relativistic heavy ion collisions. 
The transverse momentum spectrum of 
$(K^{\star0}+\overline{K}^{\star0})/2$ 
from central Au+Au collisions at $\sqrt{s_{_{NN}}}=130$ GeV is presented. 
The ratios of the $K^{\star0}$ yield derived from these data to the 
yields of negative hadrons, charged kaons, and $\phi$ mesons have been 
measured in central and minimum bias collisions and compared 
with model predictions and comparable $e^{+}e^{-}$, $pp$, and $\bar{p}p$ 
results. The data indicate no dramatic reduction of $K^{\star0}$ production 
in relativistic heavy ion collisions despite expected losses 
due to rescattering effects. 
\end{abstract}
\maketitle

  Modification of meson resonance production rates and their 
  in-medium properties are among the proposed signals of a possible 
  phase transition of nuclear matter to a deconfined plasma of quarks 
  and gluons in relativistic heavy ion collisions~\cite{rapp1}.  
  For resonances like ${K^{\star0}}$ with a lifetime comparable to the time 
  scale for evolution of the dense matter created in such 
  collisions, characteristic properties such as 
  width, branching ratio, yield, and transverse momentum spectra 
  are expected to be sensitive to the dynamics and chiral 
  properties of the high energy density medium which is 
  produced~\cite{rapp1,schaffner}. 

  More generally, the study of short-lived hadronic 
  resonances as a means to utilize the extended spectrum of 
  hadronic matter to probe hadron production under extreme 
  conditions also provides important insight into the relative 
  probability that a quark-antiquark pair will form a vector 
  resonance meson as compared to it pseudoscalar partner~\cite{derrick}. 
  This relates directly to the role of spin in hadron production in 
  strongly interacting matter under extreme conditions. 
  Additionally, the study of higher level resonances 
  affords a better understanding of feed-down to stable particles 
  from resonance decays~\cite{lep} to further constrain thermal models 
  of particle production~\cite{nxu,becattini,pbm} in nucleus-nucleus 
  collisions. 

  Resonances which decay into strongly interacting 
  hadrons in the dense matter are less likely to be reconstructed 
  due to rescattering of the daughter particles. Resonances with 
  higher $p_T$ have a larger probability of decaying outside the 
  system and therefore of being detected. 
  Alternatively, the resonance yield could be increased 
  during the rescattering phase between chemical freeze-out (vanishing 
  inelastic collisions) and kinetic freeze-out 
  (vanishing elastic collisions)~\cite{bebie,shuryak,koch} 
  via elastic processes like 
  $\pi{K}\rightarrow{K^{\star0}}\rightarrow\pi{K}$. 
  This regeneration mechanism would partially compensate for resonance decays 
  if the expansion of the produced matter took a relatively long time 
  ($^{>}_{\sim}20$  fm/$c$), increasing the observed ratio of ${K^{\star0}}/K$.  
  By systematically comparing the yields and transverse momentum distributions 
  of resonances with other particles, it should be possible to distinguish 
  different freeze-out conditions~\cite{rafelski,bravina}, such as sudden 
  freeze-out or a slow expansion of the final state hadrons. 

  Another reason the study of the $K^{\star0}$ is interesting is its 
  strange quark content. The enhancement of strangeness production in heavy ion 
  collisions has long been predicted to be a signature of the formation of 
  a deconfined quark-gluon plasma (QGP)~\cite{rafelski2}. 
  The combined measurement of the $K^{\star0}$ and $\phi$ mesons provides an additional, 
  unique tool to distinguish various hadronic expansion and freeze-out 
  scenarios~\cite{rafelski,lep,dover}. 

  The $K^{\star}(892)^{0}$ and its antiparticle are the dominant resonances in 
  the $K\pi$ system~\cite{alston}. In previous relativistic heavy ion experiments 
  the observation of these resonances has been problematic due to backgrounds from other 
  $K\pi$ partial waves~\cite{chung}, decays of higher mass resonances~\cite{pdg}, 
  elliptic flow~\cite{flow}, and detector limitations (particle misidentification, 
  acceptance and efficiency, etc.). Due to the increased energy of the beams available at 
  the Relativistic Heavy Ion Collider (RHIC), it was expected from simulation that 
  the yield of these resonances would be sufficiently large for them to be observed using 
  the mixed event method successfully used to reconstruct the $\phi$ meson at 
  RHIC~\cite{phi}. 

  The detector system used for these studies was the Solenoidal Tracker at RHIC (STAR). 
  The main tracking device within STAR is the Time Projection Chamber(TPC)~\cite{tpc} 
  which is used to provide momentum information and particle identification for charged 
  particles by measuring their ionization energy loss ($dE/dx$). 
  A minimum bias trigger was defined using coincidences between two Zero Degree Calorimeters 
  (ZDC) which measured the spectator neutrons.  
  A Central Trigger Barrel (CTB) constructed of scintillator paddles surrounding 
  the TPC was used to select small impact parameter ``central'' collisions by 
  selecting events with high charged particle multiplcity. 

  Data were taken for Au+Au collisions at $\sqrt{s_{_{NN}}}=130$ GeV. 
  To achieve uniform acceptance in the pseudorapidity range studied~\cite{hminus}, the 
  collision vertex was required to be within $\pm95$ cm of the midpoint of the TPC along the 
  beam direction. Approximately 440,000 central and 230,000 minimum bias events were used in 
  this analysis. Particles were selected based on their momenta ($p$), track quality, 
  and particle identification from the TPC $dE/dx$. Since the daughters of 
  $K^{\star0}$ decays originate at the interaction point, tracks were selected whose distance 
  of closest approach to the primary interaction vertex was less than 3.0 cm. 
  Charged kaons were selected by requiring their $dE/dx$ to be within two standard deviations 
  ($2\sigma$) of the expected mean. 
  A looser $dE/dx$ cut of $3\sigma$ was used for pions. Kaons and pions were required to have 
  transverse momenta ($p_{T}$) between 0.2 and 2 GeV/$c$ to enhance track quality~\cite{hminus}. 
  In addition, the daughters were required to have pseudorapidities $|\eta|<0.8$ with an 
  opening angle of $>0.2$~rad between them. 

  The decay channels $K^{\star0}\rightarrow\pi^{-}K^{+}$ and 
  $\overline{K}^{\star0}\rightarrow\pi^{+}K^{-}$,  both of which have a branching ratio 
  of $2/3$, were selected for the measurements. Due to limited statistics, it was 
  necessary to combine these spectra. Therefore, the $K^{\star0}$ 
  yields presented in this paper correspond to the average value of 
  $K^{\star0}$ and $\overline{K}^{\star0}$ unless otherwise specified. 
  To measure these yields, the invariant mass was calculated for each oppositely charged 
  $K\pi$ pair in an event. The invariant 
  mass distribution derived in this manner was then compared to a reference distribution 
  calculated using uncorrelated kaons and pions from different events. Typically, three or more 
  events with similar multiplicity and collision vertex locations ($|\Delta Z|<20$ cm) were 
  used for this ``mixed-event" technique. 

\begin{figure} 
  \centering 
  \includegraphics[totalheight=3in]{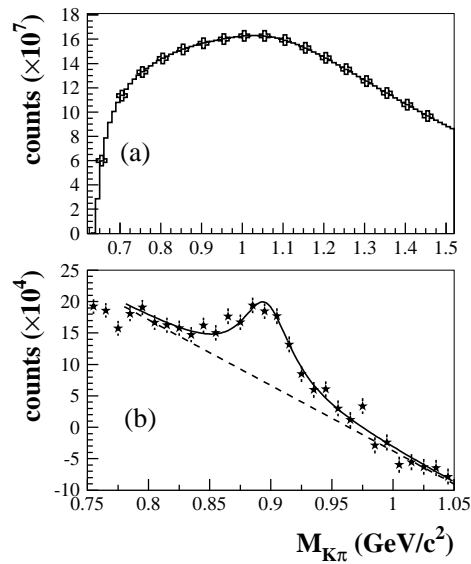} 
  \caption{a. 
  \protect{$K\pi$} invariant mass distribution from same-event pairs (symbols shown 
  every 50 MeV) and mixed-event pairs (histogram) 
  from central collisions for 0.4 GeV/\protect{$c<p_{T}< $2.8 GeV/$c$}. 
  b. \protect{$K^{\star0}$} invariant mass distribution after 
  subtraction of the mixed-event reference distribution. A Breit-Wigner 
  functional form (solid curve) was fit to the peak assuming a linear background 
  residual (dashed line). 
  The mass and width of the resonances used for the fit were fixed from 
  the Particle Data Book~\cite{pdg}. The data points reflect a bin size on 
  the x-axis of 10 MeV per bin. 
  } 
  \label{fig:kstar_mass} 
  \end{figure}

  From the 440,000 events in the 14\% most central event data sample, more than
  $1.4\times10^{10}$  
  oppositely charged kaon and pion pairs were analyzed. The corresponding invariant mass 
  distribution is shown in Fig.\ref{fig:kstar_mass}.a along with the mixed event reference 
  distribution. The two distributions were normalized to each other at 
  $M_{K\pi}\simeq$ 1 GeV/$c^2$ which is close to the mass region of interest for this measurement. 
  The two distributions are observed to match well; when subtracted the resulting distribution 
  exhibits a $K^{\star0}$  signal which is approximately 15 standard deviations above the background 
  (Fig.\ref{fig:kstar_mass}.b). The signal to background ratio before background subtraction is about 
  1/1000 for central events and 1/200 for minimum bias Au+Au interactions. These ratios are 
  significantly smaller than the value of 1/4 observed for proton-proton interactions at the CERN 
  Intersecting Storage Rings (ISR)~\cite{isr}, indicating the increased difficulty of this 
  measurement in the high multiplicity environment typical of relativistic nucleus-nucleus 
  collisions. 

  As mentioned previously, higher mass resonant states 
  in the $K\pi$ system as well as nonresonant $K\pi$ correlations also 
  contribute to the same-event spectrum. In addition, particle misidentification 
  of the decay products of the $\rho$, $\omega$, $\eta$, and $K^{0}_{s}$ cause false 
  correlations to appear in the same-event spectrum which are not present in the 
  mixed-event spectrum used to estimate the background. Comparison of the real 
  invariant mass distribution to a reference distribution derived using the 
  HIJING~\cite{hijing} event generator suggests that the residual correlation near 
  the $K^{\star0}$ mass peak may be due to the above sources. 
  However, accurate determination 
  of the magnitude of this residual correlation requires a detailed knowledge of the 
  particle production and phase space distributions for the above particles, including those 
  for the $\rho$ and $\omega$ which are presently unmeasured. Several functional forms, 
  including linear and exponential, were used to fit the residual background in 
  Fig.\ref{fig:kstar_mass}.b. The choice of normalization for the mixed-event spectrum 
  was also varied in order to study the stability of the resulting $K^{\star0}$ yield. 
  The resulting differences in yield were within 20\% in all cases, which was taken as  
  a measure, in part, of the systematic uncertainty. 

  The uncorrected number of $K^{\star0}$ was calculated 
  by integrating the Breit-Wigner function fit to the data assuming the linear residual 
  background shown in Fig.~\ref{fig:kstar_mass}.b. In order to determine the yield, 
  detector acceptance and efficiency corrections were applied as well as a correction for the 
  branching ratio. This was done by embedding simulated kaons and pions from 
  $K^{\star0},\overline{K}^{\star0}$ decays into real events using GEANT, and passing them through 
  the full reconstruction chain~\cite{hminus}. The acceptance and efficiency factor $\epsilon$ 
  depends on centrality, $p_{T}$, and the rapidity of the parent 
  and daughter particles. It varied from about 10\% for 
  parent $p_{T}\simeq0$ GeV/$c$ to approximately 35\% for parent $p_{T}\simeq2.0$ GeV/$c$. 

  \begin{figure} 
  \centering 
  \includegraphics[totalheight=2.in]{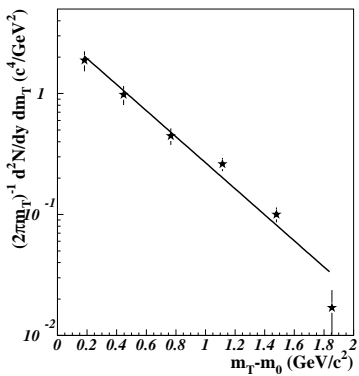} 
  \caption{\protect{The transverse mass $m_{T}$} spectrum of 
  \protect{$(K^{\star0}+\overline{K}^{\star0})/2$} within \protect{$|y|<0.5$} for the 14\% 
  most central Au+Au interactions was studied. $K^{\star0}$ resonances having  
  0.4 GeV/\protect{$c$$<p_{T}<$2.8 GeV/$c$} were detected. 
  Error bars are statistical only.} 
  \label{fig:spectra} 
  \end{figure} 
  Fig.~\ref{fig:spectra} shows $d^{2}N/(2{\pi}m_{T}dm_{T}dy)$ as function of 
  $m_{T}-m_{0}=\sqrt{p_{T}^{2}+m_{0}^{2}}-m_{0}$ where $m_0$ = 0.896 GeV/$c^2$ 
  is the mass of the $K^{\star0}$ resonance~\cite{pdg}. 
  An exponential fit was used to extract the 
  $K^{\star0}$ yield per unit of rapidity around mid-rapidity, as well as the inverse 
  slope ($T$).  The fit yielded $dN/dy=10.0\pm0.9$(stat) and $T=0.40\pm0.02$(stat) GeV 
  for central collisions. 
  The systematic uncertainty in $dN/dy$ and $T$ was estimated to be 25\% and 10\% 
  respectively due to uncertainty in the tracking efficiency and in the determination 
  of the background. Due to limited statistics, the inverse slope parameter derived for the 
  central event sample was also used to extract $dN/dy$ for the minimum bias sample. The 
  result is $4.5\pm0.7$(stat)$\pm1.4$(sys). The additional systematic error in this instance 
  results from an estimate of the uncertainty in the inverse slope of the $m_T$ 
  spectrum. Combining all $p_T$ bins, separate mass spectra of $K^{\star0}$ and 
  $\overline{K}^{\star0}$  were also fitted with a Breit-Wigner 
  resonant function plus a linear residual background. The ratio of 
  $\overline{K}^{\star0}/K^{\star0}=0.92\pm0.14$(stat) was obtained for central events. 
  Consequently, the average of the combined $K^{\star0}$ and $\overline{K}^{\star0}$ 
  spectra should accurately represent $K^{\star}(892)^{0}$ production within our statistics. 
  This ratio is similar to $K^{-}/K^{+}$ ratio~\cite{kaon}.

  The $K^{\star0}/h^-$ ratio for the top 14\% most central collisions is 
  $0.042\pm0.004$(stat)$\pm0.01$(sys) and $0.059\pm0.008$(stat)$\pm0.019$(sys) for minimum 
  bias collisions where $h^{-}$ is the total negative hadron yield 
  with $|\eta|<0.5$~\cite{hminus}. These results can be compared with 
  $K^{\star0}/h^{-}=0.036\pm0.002$ from $e^{+}e^{-}$ collisions~\cite{pdg,lep,sld} at 
  $\sqrt{s}=91$ GeV and $K^{\star0}/\pi^{-}=0.057\pm0.009\pm0.009$ from $pp$ 
  collisions~\cite{isr} at $\sqrt{s}=63$ GeV. The ratio $K^{\star0}/h^{-}$ is observed to be 
  approximately constant from low to high multiplicities at RHIC and is compatible 
  with that measured in elementary particle collisions ($e^{+}e^{-}$, $pp$, and $\bar{p}p$). 

\begin{figure} 
  \centering 
  \includegraphics[totalheight=2.5in]{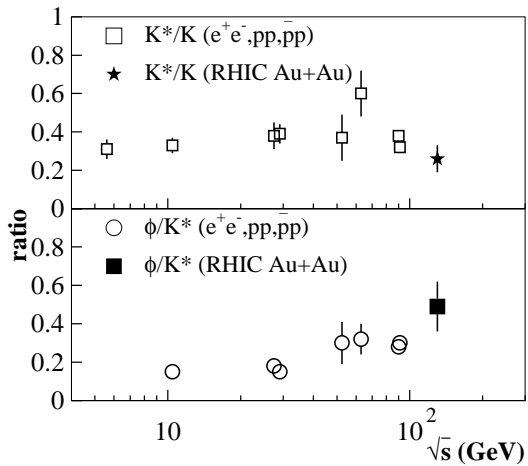} 
  \caption{The \protect{$K^{\star0}$} to charged kaon and \protect{$\phi$} to 
  \protect{$K^{\star0}$} ratios for different colliding 
  systems as a function of \protect{$\sqrt{s}$}. Data are shown with quadratically 
  combined systematical and statistical errors. 
  The data are from collisions of 
  \protect{$e^{+}e^{-}$} at \protect{$\sqrt{s}$} of 10.45  GeV, 29 GeV and 
  91 GeV~\cite{argus,derrick,lep,sld}, \protect{$\bar{p}p$} at 5.6 GeV~\cite{canter}, 
  and $pp$ from the ISR~\cite{isr,drijard} at 63 and 52.5 GeV and NA27~\cite{na27} at 28 GeV. 
  Ratios shown are for total integrated yields except for the present measurements ($|y|<0.5$) 
  and those from the ISR~\cite{isr} at \protect{$\sqrt{s}$}=63 GeV where the ratio is for 
  midrapidity.} 
  \label{fig:sqrts} 
  \end{figure}

  With respect to studies of freeze-out conditions, the ratio $K^{\star0}/K$ is more interesting 
  and less model dependent than $K^{\star0}/h^-$  since both particles 
  have similar quark content and differ only in their spin and mass. By simple spin statistics, 
  the vector meson to meson (pseudoscalar+vector) ratio would be $0.75$. However, 
the measured ratio is much smaller in elementary collisions~\cite{lep}. The charged kaon results 
  used here are an average of $K^{+}$ and $K^{-}$ in the same centrality range 
  from Ref.~\cite{kaon}. The result, $K^{\star0}/K=0.26\pm0.03$(stat)$\pm0.07$(sys) in 
  central Au+Au collisions at RHIC, can be compared with the average value of $0.37\pm0.02$ from 
  $e^{+}e^{-}$~\cite{argus,derrick,lep,sld}, $\bar{p}p$~\cite{canter} and 
  $pp$~\cite{isr,drijard,na27} as shown in Fig.~\ref{fig:sqrts}. 

  In elementary collisions, the $\phi/K^{\star0}$ ratio measures the strangeness 
  suppression to good approximation since there is only a small mass 
  difference between the $\phi$ and the $K^{\star0}$, but the strangeness quantum number 
  for these particles differs by one unit, strangeness being hidden 
  in the $\phi$ meson. This ratio seems to increase in elementary particle collisions as 
  a function of center-of-mass energy($\sqrt{s}$). In this study, it was found that 
  $\phi/K^{\star0}=0.49\pm0.05$(stat)$\pm0.12$(sys) for the 14\% most central 
  collisions~\cite{phi}. This result is greater than that from elementary collisions as 
  shown in Fig.~\ref{fig:sqrts}. The increased ratio at RHIC, compared to that in 
  elementary processes, may be indicative of strangeness enhancement and/or 
  additional effects ({\it e.g.} rescattering, coalescence~\cite{dover}) on resonances in 
  heavy ion collisions. 

  Since the lifetime of $K^{\star0}$ is comparable to the time scale for evolution of 
  the Au+Au collision system, the $K^{\star0}$ survival probability must be accounted for 
  when comparing the results from Au+Au collisions with those from elementary particle collisions 
  or with thermal model fits at chemical freeze-out. In general, 
  the $K^{\star0}$ survival probability depends on the duration of the source ($\Delta t$), 
  the size of the source for particle emission, and the $p_{T}$ of the parent $K^{\star0}$.  
  If it is assumed that the difference between the $K^{\star0}/K$ ratio in heavy ion 
  collisions and that observed in collisions of simpler systems is due to this survival probability alone, 
  the indication would be that $\Delta t$ ($^{<}_{\sim}4$ fm/$c$) is small. 
  For large $\Delta t$ ($^{>}_{\sim}20$ fm/$c$)~\cite{bravina} without $K^{\star0}$ 
  regeneration, the $K^{\star0}$ production should be an order of magnitude lower than 
  the observed result~\cite{rafelski}, and the low $p_{T}$ part of the transverse momentum 
  distribution should be suppressed resulting in a larger effective inverse $m_{T}$ slope. 
  Although the measured $K^{\star0}$ inverse slope is larger than that of the charged kaons 
  (most likely the result of radial flow~\cite{kaon,nxu}), it is still similar to that for the  $\phi$ 
  ($T=379\pm50$(stat)$\pm45$(sys) MeV)~\cite{phi}. This is consistent with the interpretation of a 
  short time (small $\Delta t$) between chemical and kinetic freeze-out. 

  Alternatively, elastic processes such as 
  ${\pi}K\rightarrow K^{\star0} \rightarrow {\pi}K$ are operative between chemical 
  and kinetic freeze-out and partially regenerate 
  the $K^{\star0}$ until kinetic freeze-out. In a  statistical model 
  description~\cite{nxu,becattini,pbm}, the measured $K^{\star0}$ should reflect 
  conditions at kinetic freeze-out rather than at chemical freeze-out if there is a long lived 
  phase in which significant rescattering takes place. 
  Within the framework of this type of model, reasonable values~\cite{nxu,becattini,pbm} 
  of the chemical and kinetic freeze-out temperatures $T_{ch}$ and $T_{th}$, 
  pion chemical potentials $\mu_{\pi}$ at kinetic freeze-out~\cite{rapp1,shuryak}, and the mass 
  difference ($\Delta m = 0.4$ GeV) between $K^{\star0}$ and $K$ can be obtained. 
  These result in the ratio of $(K^{\star0}/K)_{th}$ at kinetic freeze-out and 
  $(K^{\star0}/K)_{ch}$ at chemical freeze-out being roughly 
  $(K^{\star0}/K)_{th}/(K^{\star0}/K)_{ch}=\exp{[(-\Delta m+\mu_{\pi})/T_{th}+\Delta
  m/T_{ch}]}$ 
  which is in the range of 0.3 to 1.2. The value measured in the present study ($0.26$) for kinetic 
  freeze-out and the value of $0.37$ assumed for chemical freeze-out are in the ratio of $0.7\pm0.2$, 
  which selects specific trajectories in the phase diagram of $\mu_{\pi}$/$\mu_B$ and $T$~\cite{shuryak}.
  The ratio at chemical freeze-out is 
  based on a statistical model~\cite{nxu,becattini,pbm}, and upon collisions of lighter systems where 
  the chemical and kinetic freeze-out processes are naturally overlapped. 

  Present measurements are not consistent with a long expansion time 
($\Delta t^{>}_{\sim}20$ fm/$c$) without significant $K^{\star0}$ regeneration. They are consistent 
  with a sudden freeze-out interpretation ($\Delta t^{<}_{\sim}4$ fm/$c$). 
  The simple estimates made in this paper illustrate how the study of resonance 
  can provide important information on the dynamics and evolution of the 
  matter produced in relativistic nucleus-nucleus collisions. More sophisticated analyses 
  (e.g. comparison with transport models such as UrQMD~\cite{bleicher}) with improved 
  uncertainty in the measurement of the $\phi$ and $K^{\star0}$, 
  as well as the measurement of additional resonances are needed to determine 
  the evolution of the system resulting from central heavy ion collisions in detail. 

  In conclusion, we have presented the first measurement of $K^{\star}(892)^{0}$ 
  and $\overline{K}^{\star}(892)^{0}$ in relativistic nucleus-nucleus collisions. 
  The $K^{\star0}$ $m_{T}$ spectrum from the 14\% most central Au+Au collisions 
  results in an inverse slope parameter similar to that measured for the $\phi$ meson 
  in similar centrality. The measured yield, 
  $dN/dy=10.0\pm0.9$(stat) $\pm2.5$(sys) is relatively high compared to elementary 
  collisions and thermal model predictions, considering the short $K^{\star0}$ lifetime 
  ($c\tau\simeq4$ fm) and expected losses due to rescattering of the decay daughters in 
  the dense medium. The results of this study are consistent with two possible scenarios 
  for the dynamic evolution of the system: 
  (1) a short time duration between chemical and kinetic freeze-out ({\it i.e.} sudden 
  freeze-out), or (2) a long period of expansion characterized by high hadron density 
  and significant $K^{\star0}$ regeneration along specific trajectories in the phase 
  diagram. Studies of strongly decaying resonant states 
  like the $K^{\star0}$ open a new approach to the study of relativistic nucleus-nucleus 
  collisions. 

We wish to thank the RHIC Operations Group and the RHIC Computing Facility
at Brookhaven National Laboratory, and the National Energy Research 
Scientific Computing Center at Lawrence Berkeley National Laboratory
for their support. This work was supported by the Division of Nuclear 
Physics and the Division of High Energy Physics of the Office of Science of 
the U.S. Department of Energy, the United States National Science Foundation,
the Bundesministerium fuer Bildung und Forschung of Germany,
the Institut National de la Physique Nucleaire et de la Physique 
des Particules of France, the United Kingdom Engineering and Physical 
Sciences Research Council, Fundacao de Amparo a Pesquisa do Estado de Sao 
Paulo, Brazil, the Russian Ministry of Science and Technology and the
Ministry of Education of China and the National Natural Science Foundation 
of China.

\end{document}